\documentclass[showpacs, showkeys, 
twocolumn, 
amssymb, amsmath,
nobibnotes,aps,prd]{revtex4}
\usepackage{amsmath}
\usepackage{graphicx}
\begin{document}
\title{Strong cosmic censorship: the role of nearly extreme nonrotating black holes}
\author{Wenceslao \surname{Santiago-Germ\'{a}n}}
\affiliation{Department of Physics, The University of California, Davis, CA 95616-8677, USA}
\email{wenceslao@lifshitz.physics.ucdavis.edu}
\date{March 22, 2004}
\begin{abstract}
 In this paper it is argued that even invoking the dominant energy condition, the Einstein's field equations admit nonglobally hyperbolic, asymptotically de Sitter spacetimes with locally naked singularities and with Cauchy horizons failing to be surfaces of infinite gravitational blueshift; what is more, it is shown the existence of space-time solutions containing Cauchy horizons expected to be stable against small linear nonstationary axisymmetric perturbations in an open set of the physical parameter space where the  positive cosmological constant $\Lambda$ is nearly zero. The price to pay is a violation at ultrahigh densities of Le Chatelier's principle. Given an equation of state, the line of thought introduces black-hole Cauchy horizon instability criteria in terms of the associated adiabatic index $\Gamma,$ revealing a connection between this subject and the stability theory of relativistic stars. This is used to put forward a tentative, limited formulation of the strong cosmic censorship principle inside black holes \textemdash once called dark stars.   
\end{abstract}
\pacs{04.20.Dw, 04.70.Bw}  \keywords{Black holes, singularities} 
\maketitle

\section{\label{sec:intro}Introduction}

 In 1969, Penrose \cite{Penrose1969} summarized what was known  at the time on the subject of gravitational collapse and  presented a coherent picture. In his paper, ``Gravitational Collapse: the Role of General Relativity," he raised  the  question that remains arguably the most outstanding unsolved mathematical problem of the field: ``Is there a cosmic censor, who forbids the appearance of naked singularities, clothing each one in an absolute event horizon?"  
 
  According to the singularity theorems of Hawking and Penrose \cite{PenroseST,Tip,Ellis}, the Einstein's law of gravitation predicts, under certain fairly physical situations of unstoppable gravitational collapse, that space-time singularities, i.e., causal geodesic incompleteness, must occur. The methods of proof  leading to this remarkable conclusion were, however, incapable of revealing the  structure of the geometrical breakdown. And in \cite{Penrose1969}, it was conjecture that:   
  
  \textit{The gravitational collapse of an isolated `physical' body, starting from generic regular initial data, cannot produce spacetime singularities that can be seen from infinity, even though observations from infinity are allowed to continue indefinitely.}
  
   Thus, the `physical' singularities formed by the gravitational collapse of astrophysical bodies are conjectured to be  always hidden inside black holes. 
   
   It should be pointed out that this claim, now referred to as the \textit{weak cosmic censorship conjecture,} has been used as theoretical assumption in several important results; for instance, it appears as an hypothesis in  Hawking's area theorem \cite{Hawking}, which led to the discovery of a basic connection between  the mechanics of black holes and the laws of thermodynamics \cite{Bardeen}.  
  
\begin{figure}[t] 
\includegraphics[width=3in]{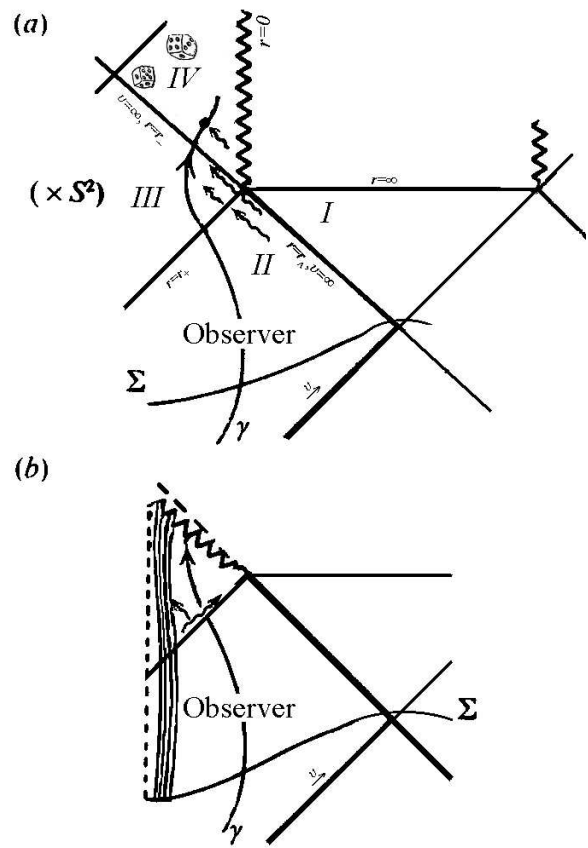} 
\caption{\label{wormhole} A piece of the conformal Carter-Penrose diagram of a charged black hole in an asymptotically de Sitter spacetime: (a) static spherically symmetric with a locally naked singularity \textemdash also depicted, an ingoing flux of radiation falling inside the hole; (b) ``generically" perturbed around a particular state where the black Cauchy horizon is unstable, also shown collapsing matter and the scattering of an outgoing mode. At $r_-$ we have the inner Cauchy horizon; i.e the boundary of the region that can be predicted from knowledge of initial data given in $\Sigma$; at $r_+$ the absolute event horizon; at $r_{\Lambda}$ the cosmological horizon; thick wiggle lines for space-time singularities; curve $\gamma$ for the path of an observer. The physics in region IV is determined by information coming not only from $\Sigma$ but also from a locally naked singularity whose behavior is unknown.}
\end{figure}

  Weak cosmic censorship,  nevertheless, allows for singularities that are locally visible, although not accessible from observation from infinity. A  singularity in the structure of the spacetime is said to be \textit{locally naked} if it lies both to the past of a nonsingular spacetime event and to the future of some other regular point of the spacetime continuum, see Fig\ref{wormhole}(a). Thus, according to the first part of the definition, in principle  uncontrollable information may enter the spacetime from a locally naked singularity upsetting the unique predictability of the outcomes of observation and spoiling the uniqueness in the large of the space-time continuum. That possibility alone \textemdash and an intuitive picture about the stability of the internal structure of charged black holes (see the rest of the introduction), led Penrose \cite{Penrose1978, Penrose1979} to  conjecture  that generically the classical general theory of relativity would not allow for the creation of  locally naked space-time singularities; in other words:   
  
  \textit{``Every inextendible spacetime $\mathcal{M}$ which evolves, according to classical general relativity with physically reasonable matter satisfying appropriate energy conditions, from generic non-singular initial data on a complete spacelike hypersurface $\Sigma,$ is globally hyperbolic."} 
  
  The former statement, now referred to as the \textit{strong cosmic censorship hypothesis,} implies that to examine a `physical' space-time singularity, an observer must run into it, see Fig\ref{wormhole}(a); the Big Bang singularity does not count as locally naked. A review of the subject can be found in references \cite{Chrusciel,Penrose1998}.
  
Cosmic censorship research consists in finding  an exact  mathematical formulation of the above conjectures and providing a rigorous proof of the proposed statement.   

The subtle terms: `generic gravitational collapse' \cite{Christ1999}, `physically reasonable matter' \cite{Yodzis}, and `appropriate energy conditions' \cite{Penrose1998} are essential to the validity and the proper formulation of the cosmic censorship hypothesis since instances are known where, in theory, naked singularities do form dynamically from regular initial data \cite{Yodzis, Smarr}. The claim, however, is that  spacetime singularities which are not `hidden from view'  must be physically impossible to achieve; say, because they are unstable or because the stable ones can only arise from matter fields which do not exist in nature. Unfortunately, the actual knowledge of the matter content of the universe is incomplete \cite{Riess} and  the global nonlinear stability theory of Einstein's field equations remains almost unexplored \cite{milenium}. Thus, serious difficulties would have to be overcome to formulate and verify the cosmic censorship conjecture. 

To avoid dealing with the  details and complications of an unknown energy momentum tensor, one usually resorts to the use of energy conditions. One of these is the dominant energy condition stating that: For every timelike vector $W_\mu,$ the stress-energy tensor satisfies $T^{\mu \nu}W_\mu W_\nu \geqslant 0$, and $T^{\mu \nu}W_\nu$ is a nonspacelike vector. It implies that matter cannot travel faster than light \cite{Ellis}. 

 What is, one may ask, the theoretically precise  meaning of the qualifications: `generic', `physically reasonable' matter, and `appropriated' energy conditions; how do they fit together in a mathematical description of the inner workings of, say, the cosmic censor described by the second riddle: the strong cosmic censorship? Nobody knows for sure. 
 
   The purpose of this paper is to gain some intuition, and possibly a hint, about this problem  by using the linear perturbation theory of black-hole interiors: carrying a step further Penrose's original arguments \cite{Penrose1979}. Specifically, we invoke black holes with multiple horizons: inner Cauchy, absolute, and cosmological horizons \textemdash see Fig.\ref{wormhole}(a), with their own surface gravities:  $\kappa_-, \kappa_+,$ and  $\kappa_\Lambda$ respectively, which play the role of temperature in the laws of black hole mechanics \cite{Bardeen}. 
     
The present article follows a line of reasoning that owns its origins to an observation by Simpson and Penrose \cite{Simpson}, who noted that,  presumably, the inner structure of the analytically extended  Reissner-Nordstr\"om background, along with its locally naked singularities, is unstable under perturbations that break spherical symmetry. The reason: the infinite amount of gravitational blueshift, on the inner Cauchy horizon, for the flux of radiation that has fallen into the hole and derived from the smallest source in the `outside world': Radiation made of gravity and electromagnetic waves, carrying angular momentum and imparting (in general) a small rotation to the hole. This intuition was confirmed mathematically by  subsequent analytical studies of the underlying black hole perturbation theory not only in the asymptotically flat case \cite{McNamara, Chandra} but also in the asymptotically de Sitter scenario \cite{Moss}.
  
  To quantitatively describe  the situation, consider for a moment Fig.\ref{wormhole}(a). Using ingoing Eddington-Finkelstein coordinates, the geometry in regions $II$ and $III$  can be described by the line element $ds^2= -e^{2\psi}Fdv^2 + 2e^{\psi}drdv+r^2(d\theta^{2}+\sin \theta^2 d\phi^2).$  Now define the radial null coordinate $v$ as an advance time in region $II,$ so that $r$ decreases towards the future  along a ray $v=const.$ More precisely, in terms of the tortoise coordinate $r_\ast$, satisfying  $dr_\ast=\int (e^{\Psi}F)^{-1} dr,$ let it be given by the relationship: $\upsilon=r_{\ast}+t$ for $r_+<r<r_{\Lambda},$ where $\partial/\partial t$ is the time-translation killing vector field with the normalization $g^{\mu\nu}t_{,\mu} t_{,\nu}=-1$ at asymptotically flat infinity when $\Lambda$ is set to zero. It is then analytically continued for the remaining values of $r$.

The formulae quantifying the lowest degree of divergence of the flux of radiation $\mathcal{F}$  (modeled as a linear perturbation) incident on an observer at the instant of his crossing the Cauchy horizon, precisely at the moment when the advanced time $\upsilon$ tends to  $+\infty$  (see Fig.\ref{wormhole}(a)) is given by     
\begin{equation} 
\mathcal{F} \rightarrow  constant  \;\; e^{(\kappa_--\kappa_+)\upsilon}  \label{Funiv}
\end{equation}
for an asymptotically flat universe under perturbations of \textit{compact} support \cite{Chandra} and  
\begin{equation}
\mathcal{F} \rightarrow   constant \;\; e^{\kappa_{-} v} \left(e^{-\kappa_{\Lambda} v} +  constant \times e^{-\kappa_{+} v}\right)  \label{Duniv}
\end{equation} 
for an asymptotically de Sitter universe \cite{Moss}. 

Heuristically, these results can be understood as follows. Take the asymptotically de Sitter case; Eq.(\ref{Funiv}) can be obtained by carefully taking the limit when $\Lambda$ goes to zero while keeping the perturbation compact. The influx of radiation along the Cauchy horizon is produced by the transmission of ingoing modes coming from the asymptotic region \textit{close} to the cosmological horizon, see Fig.\ref{wormhole}(a). This radiation is gravitationally redshifted by climbing away from the cosmological horizon and blueshifted by falling into the inner horizon; thus the $e^{-\kappa_\Lambda v}$ and $e^{\kappa_- v}$ factors respectively. However, there is also an additional influx due to the scattering, by the black hole curvature, of outgoing modes originated in the \textit{neighborhood} of the event horizon, see Fig.\ref{wormhole}(b). Since this contribution comes from a region of different gravitational potential, the    redshift is controlled now by $e^{-\kappa_+ v},$ whereas the blueshift is a function of $e^{\kappa_-v}$ as before.
    
It is worth pointing out that the internal structure of a black hole depends of the fate of the universe itself, and that the proportionality constants appearing in (\ref{Funiv}) and (\ref{Duniv}) contain the relevant information about such a destiny. Both the topology of the space and the type of perturbations considered on the black hole background determine whether or not such constants yield finite or infinite numbers \cite{Penrose1979, Chandra, Moss}.  
  
Whereas for a Reissner-Nordstr\"om--de Sitter spacetime $\kappa_-$ is always larger than $\kappa_+$ provided that  $r_-\neq r_+$ (hence the divergence of $\mathcal{F}$ on the Cauchy horizon \cite{Moss}), and since such a hole satisfies the dominant energy condition, the novelty of our result consist in showing that: Even invoking the dominant energy condition over other fields, or combination of fields, it is still possible to make $\kappa_-$ small enough, so that, properly speaking, it becomes the smallest of the triplet: $\kappa_-$, $\kappa_+,$ and $\kappa_\Lambda.$ Thus rendering  a situation where, in principle, the Cauchy horizon, starting at a locally naked singularity, fails to be a surface of infinite gravitational blueshift and  remains stable against small linear nonstationary axisymmetric perturbations.

The outline of the paper is as follows: in section II, a connection between the value of the adiabatic index and the blueshift instability of black-hole Cauchy horizons is established. It is followed by an example, in section III, where both the dominant energy condition and the blueshift stability criteria are satisfied. This example relies, however, on a violation of Le Chatelier's principle \cite{Chatelier} in the ultrahigh density regime. We conclude with a tentative, limited formulation of the strong cosmic censorship hypothesis ruling the interior of black holes, expected to hold under nearly extreme conditions.  
 
\textit{Conventions:} Greek indices refer to the complete spacetime, Latin indices are reserved for two-dimensional quantities. Units are used where Newton's gravitational constant $G$ satisfies $8\pi G=1$ unless otherwise indicated. The constant $c$ denotes the speed of light in the vacuum. 

\section{\label{sec:reason}Line of reasoning}
 From what we have seen so far, it is clear that great insight will be gained if we manage to connect the purely geometrical expressions given by $\exp[(\kappa_- - \kappa_+)\upsilon]$ and $\exp[(\kappa_- - \kappa_\Lambda)\upsilon]$ to the stress energy content of the spacetime. It is to this study that this section is devoted.

According to the zero law \cite{Bardeen, Racz}, any static black-hole, not necessarily a solution of Einstein's field equations, has  a constant surface gravity over its event horizon; if Einstein's field equations and the dominant energy condition are satisfied; then, the surface gravity must be constant on any killing horizon.  

Let us introduce coordinates $\epsilon_i$ of points in the physical parameter space of a stationary black hole solution: \begin{equation}
\overrightarrow{\epsilon}=(\epsilon_M,\epsilon_Q,\epsilon_J,\epsilon_\Lambda,...)=(M,Q,J,\Lambda,...),
\end{equation}
where $M,Q,$ and $J$ denote the total mass, charge, and angular momentum of the hole \textemdash which may or may not determine uniquely its properties, and $\Lambda$ is a nonegative cosmological constant assumed to have a very small value.  

Since, in the physical parameter space, extreme black holes define a `boundary'  beyond which black holes do not exist; the points at `the other side' typically represent naked singularities, it is worthwhile exploring what happens in the neighborhood of such a boundary. Historically, cosmic censorship has been tested using black holes under nearly extreme conditions, most notably by Wald \cite{Wald}, without ever contradicting the spirit of the hypothesis \cite{Brill}. 

 \subsection{Nearly extreme static black holes}
  Let me direct your attention to the metric for the static spherically symmetric gravitational field produced by a spherically symmetric body  at rest,  which provides, at late times, a rough description of the geometry of one of the conceivable final states in the gravitational collapse of a  body (independently of its composition) that remains nearly spherically symmetric during its evolution. It reads
\begin{equation}
ds^2=g_{ab}dx^a dx^b + r^2d\Omega^2, \label{smetric}
\end{equation}
where $g_{ab}$ denotes a 2-dimensional Lorenztian metric, $r$  the area radius function, and $d\Omega$ the line element of the unit 2-sphere: Is there a counterexample to strong cosmic censorship in this restricted class of solutions? How can we find a counterexample?

  Of special interest is to investigate the blueshift instability question of black-hole Cauchy horizons associated to static space-times of the form (\ref{smetric}). To this end, it is convenient to work with the null coordinate of the advanced time $\upsilon,$ and the following functions of $r:$
\begin{equation}
F(r) \equiv g^{\mu \nu}r_{,\mu}r_{,\nu}, \;\;  e^{-\psi} \equiv g^{\mu \nu}r_{,\mu} \upsilon_{,\nu}, \;\; and \;\; U(r)=e^{\psi}F,
\end{equation}
demanding that at asymptotically flat infinity, when $\Lambda$ is set to zero, $F(r)|_{r \rightarrow \infty}=1$  and   $\psi(r)|_{r \rightarrow \infty}=0.$  

The horizons are located at constant values of $r;$ therefore the vector field $g^{\mu \nu}r_{,\nu}$ is normal to these hypersurfaces. Since they are null surfaces, it follows that both $F$ and $U$ have vanishing values at the black hole horizons. Let us consider also the null vectors $\ell_{\mu}=e^\psi \upsilon_{,\mu}$ and $n_{\mu}$ so that   
\begin{equation}
\ell^{\mu}\ell_{\mu}=0, \;\; \ell^{\mu}n_{\mu}=-1, \;\; and  \;\; n^{\mu}n_{\mu}=0. 
\end{equation}
From them, we derive the following stress-energy tensor
\begin{eqnarray}
T_{\mu \nu}&=&(\rho + p_{\theta})(\ell_{\mu}n_{\nu} + n_{\mu}\ell_{\nu}) \nonumber \\
 && + F^{-1} (\rho+p_r)r_{,\mu}r_{,\nu}+(p_{\theta}-\Lambda c^4) g_{\mu \nu} \label{stress}
\end{eqnarray} 
where $p_\theta$  and $p_r$  are, respectively, tangential and radial pressures, $\rho+\Lambda c^4$ is the energy density as measured by a stationary observer with four velocity $$S^{\mu}= |F|^{-1/2}(r^{,\mu}-F \ell^{\mu})$$ outside the hole. 

In terms of the mass function $m(r),$ defined by 
\begin{equation}
 F\equiv 1-(2Gm/c^{2}r), \label{mass}
\end{equation}
the Einstein's law of gravitation leads to \cite{Visser}
 \begin{eqnarray} 
G_{r \upsilon}:& \rho+\Lambda c^4=&\frac{dmc^2}{d\mathcal{V}}; \label{Erho}\\
G_{r r}:& F^{-1}(\rho+ p_r)=& 16 \pi \frac{d\psi c^4}{d\mathcal{A}} ; \label{reg}\\
G_{\theta \theta}:& p_\theta-\Lambda c^4 =&- \frac{d(\mathcal{A}\frac{dmc^2}{d\mathcal{V}})}{d\mathcal{A}} +  \frac{3}{2} F'\psi'c^{4}  \nonumber \\  && +  F(\psi^{''}+{\psi^{'}}^2 + r^{-1}\psi')c^{4}, \label{Gpressure}
\end{eqnarray}
in units where Newton's gravitational constant $G$ satisfies $8\pi G=1$; the prime indicates `partial' differentiation with respect to 
the $r$-coordinate;  $\mathcal{A}$ is $4\pi r^2$ and $\mathcal{V}$ is $ (4/3)\pi r^3.$ We say `partial', not `total', because we let  $F$ and $\psi$ to be also functions of parameters like $M,Q,J$ and $\Lambda.$

In terms of the $U$-function, the surface gravity takes a very simple form 
\begin{equation}
\kappa=|U'/2|, \label{sgdef}
\end{equation}
the radial null geodesic equation becomes  $dr/dt=\pm U(r)$ in `Schwarzschild-like' coordinates  \footnote{Note that $U^{-1}$ corresponds to the $\varphi$ function of \cite{Giambo}}, and the four-dimensional wave equation $$\Box e^{i(m\phi-\omega t)} P^{m}_{\ell}(\cos \theta) r^{-1}Z(r) =0,$$ reduces to 
\begin{eqnarray}
(\frac{d^2}{dr_{\ast}^2} + \omega^2)Z&=&VZ;  \\ 
V &=& U(U'r^{-1}+  \ell(\ell+1) e^{\psi}r^{-2}), 
\end{eqnarray}
where $P^m_{\ell}(z)$ is the associated Legendre polynomial.  The $\ell(\ell+1) e^{\psi}r^{-2}$ part of the effective potential $V$  is a centrifugal barrier, the $U'$ part is due to the curvature of the spacetime: A black hole, when perturbed, can be visualized in terms of potential barriers that scatter or reflect a fraction on any incident wave, including gravity waves.   
 
 The global geometry of the intersection of the hypersuface $z=U(\epsilon_i,r)$ with the $z=0$ plane  determines the relationship among the area of each horizon with the mass, charge, and angular momentum of the hole, a dependence which in principle could be very complex, perhaps given in terms of transcendental functions. In what follows, we obtain a strictly local determination of this functional relationships  assuming the  local existence of only two solutions: $r_-=f_-(\epsilon_i)$ and $r_+=f_+(\epsilon_i)$ (respectively, the area radius of the inner and absolute horizons), in the neighborhood of an extreme state where the equality $r_+=r_-$ holds. 
 
\begin{figure}[h]
\includegraphics[width=2in]{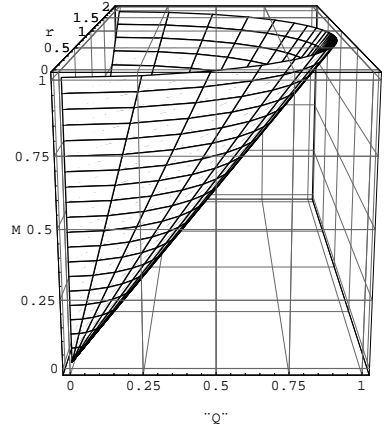} 
\caption{\label{uzero}  A pictorial representation, in $\Re^+\times\Re^2,$ of the hypersurface defined by the solutions of the equation $U(r,M,Q)=0 $ in the case of a Reissner-Nordstr\"om black hole. In this instance we have $U(r,Q,M)=1-2Mr^{-1}+Q^2 r^{-2}.$ It follows that the hypersurface shown in the figure intersects the plane of symmetry $Q=0$ only at the Schwarzschild radius. The functions $r_+=f_+(\epsilon_i)$ and $r_-=f_-(\epsilon_i)$ can be calculated explicitly, $f_{\pm}(M,Q)=M\pm (M^2-Q^2)^{1/2},$ allowing a consistency check of equations (\ref{area}) and (\ref{sg}). The curve $(r,M,Q)=(M_o,M_o,\pm M_o)$ defined in terms of the parameter $M_o\in \Re^+$ gives the extreme states of the hole.}
\end{figure}
 The following method is similar to the one given for the proof of the implicit function theorem found in \cite{Courant}.   
Suppose, for the sake of argument, that $U(r,\epsilon_i)$ is a $C^3$--differentiable function in all its variables and that $U \sim 1-\frac{2GM}{c^2r}-\frac{\Lambda r^2}{3}$ at infinite values of the $r$-coordinate. Denote also the geometric locus of points corresponding to the static, spherically symmetric, extremal black holes of the corresponding theory by $\Upsilon$; that is 
\begin{equation}
 \Upsilon \equiv \left \{  (r_o, \overrightarrow{\epsilon}_o)  | U(r_o,{\epsilon_o}_i)=U'(r_o,{\epsilon_o}_i)=0; r_o\neq r_{\Lambda} \right \}, \label{extremeSET}
\end{equation} 
where $r_\Lambda=f_{\Lambda}({\epsilon_o}_i)$ is the area radius of the cosmological horizon (if any). Therefore, on $\Upsilon,$ the $\epsilon_i$'s are not independent parameters since they have to solve the relationships given in (\ref{extremeSET}). 

In view of the smoothness and asymptotic limit of the $U$-function, the requirement that the area of the event horizon be different from the area of the cosmological horizon, that is $r_+\neq r_\Lambda,$ implies that $U''\geq 0$ on $\Upsilon$ (recall that $U'$ is obtained by fixing $\epsilon_i$ and taking the partial derivative with respect to the $r$-coordinate).  In order to fix ideas, Fig.\ref{uzero} shows a section of the intersection of the corresponding hypersurface $z=U(r,\epsilon_i)$ with the $z=0$ plane for the case of a Reissner-Nordstr\"om black hole.  

 Surely, a nearly extreme black hole is always at a very short `distance' from an extremal one, but how close are the horizon areas and the surface gravities of these black holes for a shift $\delta \epsilon_i$ at a given extreme point of the physical parameter space? Using the Taylor's series of the U-function in the neighborhood of an extreme state, a point in $\Upsilon,$ we find
\begin{eqnarray}
& \delta U(\epsilon,r)= & U_{,\epsilon_i} \delta \epsilon_i + \frac{1}{2}\frac{\partial^2 U}{\partial \epsilon_i \partial \epsilon_j} \delta \epsilon_i \delta \epsilon_j +  U^{'}_{,\epsilon_i} \delta \epsilon_i \delta r  \nonumber \\ 
& & + \frac{1}{2} U^{''}\delta r^2 + \mathcal{O}[(\delta r^2 + \delta \epsilon^2)^{3/2}],   \label{expansion}
\end{eqnarray}
and also
\begin{eqnarray}
& \delta\kappa= &  \kappa' \delta r + \kappa_{,\epsilon_i} \delta \epsilon_i + \frac{1}{2}\frac{\partial^2 \kappa}{\partial \epsilon_i \partial \epsilon_j} \delta \epsilon_i \delta \epsilon_j  +  \kappa^{'}_{,\epsilon_i} \delta \epsilon_i \delta r  \nonumber\\
& & + \frac{1}{2} \kappa^{''}\delta r^2 + \mathcal{O}[(\delta r^2 + \delta \epsilon^2)^{3/2}]. \label{surface}
\end{eqnarray}
where $\delta U=U({\epsilon_o}_i+\delta \epsilon_i, r_o+\delta r)-U({\epsilon_o}_i,r_o).$ If $\delta r=f_+({\epsilon_o}_i+\delta \epsilon_i)-r_o \equiv \delta r_+$ or $\delta r=f_-({\epsilon_o}_i+\delta \epsilon_i)-r_o \equiv \delta r_-$ we have that $\delta U$ vanishes.  Thus, if $U''> 0,$ substituting these special values in (\ref{expansion}) and solving the quadratic equation for $\delta r,$ we obtain  
\begin{equation}
\delta r_{\pm}=[\pm (-2U''U_{,\epsilon^i}\delta \epsilon_i)^{1/2} - U^{'}_{,\epsilon_i}\delta \epsilon_i](U'')^{-1}+\mathcal{O}(\delta \epsilon^{3/2}),  \label{area}
\end{equation}
where $r_-$ and $r_+$ respectively denote  the area radius of the inner and absolute horizons satisfying $\delta r_+ \geq \delta r_-,$ in the neighborhood of $\Upsilon.$  

Equation (\ref{surface}), using (\ref{sgdef}) and (\ref{area}), leads to 
\begin{equation}
2\kappa_{\pm}=(-2U''U_{,\epsilon_i}\delta \epsilon_i)^{1/2}\mp U''' U_{,\epsilon_i}\delta \epsilon_i (U'')^{-1}+\mathcal{O}(\delta \epsilon^{3/2}). \label{sg}
\end{equation}
where the partial derivatives are evaluated in the corresponding point $(r_o,{\epsilon_o}_i)$ of $\Upsilon.$
Of course, it is necessary to choose  a direction for $\delta \epsilon^i$  allowing the existence of two real roots, making the quantity inside the radical symbol nonegative.
 
Hence, in the vicinity of $\Upsilon,$ we find
\begin{equation}
e^{(\kappa_- - \kappa_+)\upsilon}= e^{\left [ (-2U''U_{,\epsilon_i}\delta \epsilon_i )\frac{(-U''')}{2(U'')^{2}}+\mathcal{O}(\delta \epsilon^{3/2}) \right ] \upsilon}. \label{rate}
\end{equation}
It is worth noting that the result given by the system of equations (\ref{area})-(\ref{rate}) is independent  of the parametrization $\epsilon_i.$ The next relevant step is to connect the purely geometrical expression on the left hand side of (\ref{rate}) to the stress energy content of the space-time.

\subsection{\label{subsec:est}Surface gravity estimates}

An inspection to the theorems of Lindblom and Masood-ul-Alam, the treatment on the subject by Shapiro and Teukolsky, and the classical work of Chandrasekhar, reveals that the relativistic adiabatic index $\Gamma$ given by
\begin{equation}
\Gamma = \frac{\rho+p}{p} \left(\frac{\partial p}{\partial \rho} \right)_S, \label{adi} 
\end{equation}
plays  a primary role in the stability theory of relativistic stars \cite{Lindblom, Shapiro, Stellar, Sri, MTW}. In the characterization of $\Gamma,$ variations of pressure and density are taken at a fixed entropy $S$ of the element of fluid; heat is neither added nor taken away. In the Newtonian limit where $\rho\gg p,$ see the discussion of Wald  \cite{WaldGR} on perfect fluids, the conventional expression taught in textbooks: $\Gamma\approx (\rho/p)(\partial p/\partial \rho)_S$ is recovered. A fundamental result of the theory due to Chandrasekhar is the following \cite{Chandra1964}: 

\textit{In the post-Newtonian approximation, where $GM/Rc^2\ll 1,$  a global adiabatic radial instability  sets in for the core of a nonrotating star of mass $M$ and radius $R$ when the pressure-average adiabatic index $\overline{\Gamma},$ governing the perturbations, satisfies the inequality $$\overline{\Gamma}<\frac{4}{3}+K\frac{GM}{Rc^2},$$ where $K$ is a constant of order unity, depending on the structure of the star, and $\Gamma$ is set equal to $(\Delta p/p)/(\Delta \rho/ \rho).$ }
 
In regard to the blueshift instability question of the interior of black holes \textemdash once called dark stars \cite{Dark}, we offer the next analysis giving a first indication of how the adiabatic index enters into the theory. 

Let us define    
\begin{equation}
\Gamma_{\|}= \frac{\rho + p_{\theta}}{p_{\theta}} \left (\frac{\partial p_\theta}{\partial \rho} \right )_S \;\; and \;\;  \Gamma_{\bot}= \frac{\rho + p_r}{p_r} \left (\frac{\partial p_r}{\partial \rho} \right )_{S}, \label{adiabatic} 
\end{equation}
governing the background pressure-density relations. 
 
 According to the system of equations (\ref{Erho})--(\ref{Gpressure}), Einstein's law of gravity implies  
\begin{equation}
p_\theta d\mathcal{A}+ d(\mathcal{A}\rho)=0    \;\; on  \;\; \Upsilon. \label{Bianchi}
\end{equation}
 Thus whenever $\Gamma_{\|}$ is constant, $\mathcal{A}^{\Gamma_{\|}}p_\theta$ is also constant on $\Upsilon$.  Note that the equality (\ref{Bianchi}) still holds if one replaces $p_{\theta}$ and $\rho$ by $p_{\theta}-\lambda c^4$ and $\rho+\lambda c^4$ respectively (with $\lambda$ constant).  Whether $\Gamma_{\|}$ is constant or not, it let us obtain, combining (\ref{Gpressure}) and (\ref{Bianchi}), a single compact formula for the third partial derivative of $U$ with respect to the area radius coordinate evaluated at a point in $\Upsilon.$
 Rewriting (\ref{Gpressure}) in the form
\begin{equation}
2(p_{\theta}-\Lambda c^4)=[\psi'' U+(2r^{-1}+\psi')U'+U'' ]e^{-\psi}c^{4}
\end{equation}
and  taking the partial derivative with respect to $r$-coordinate, we obtain
\begin{equation}
U'''e^{-\psi}r^{-2}=[(2c^{-4}p_{\theta}-2\Lambda-\psi''e^{-\psi}U)r^{-2}]' +(e^{-\psi}r^{-2})''U'.
\end{equation}
 Thus, making use of (\ref{Bianchi}) and the chain's rule, it follows
\begin{equation}
(-U''')=   4c^{-4}e^{\psi}[ (\Gamma_{\|}+1)p_{\theta}-\Lambda c^4] r^{-1} \;\;\;\; \mathrm{on} \;\; \Upsilon, \label{lcriterion}
\end{equation}
where the condition $U''>0$ is now $p_{\theta}-\Lambda c^4>0.$

Let us restrict our attention to the subset of $\Upsilon$ obtained by setting ${\epsilon_o}_\Lambda =0.$ In this case  (\ref{lcriterion}) reduces to
\begin{equation}
(-U''')=   4c^{-4}e^{\psi}(\Gamma_{\|}+1)p_{\theta} r^{-1} \;\;\;\;  on \;\; \Upsilon \cap\{{\epsilon_o}_\Lambda=0\} \label{criterion}
\end{equation}
with $p_{\theta} \neq 0.$ Imparting a shift $\delta \epsilon_i=\epsilon_i-{\epsilon_o}_i$ with ${\epsilon_o}_\Lambda=0<\Lambda=\delta \epsilon_\Lambda \ll \delta \epsilon_i$ for $i \neq \Lambda$ in (\ref{sg}) gives
\begin{equation}
2\kappa_- \approx (-2U''|_{\Lambda=0}U_{,\Lambda} \Lambda)^{1/2}.
\end{equation} 
 Moreover, since $U \sim 1 - 3^{-1}\Lambda r^2$ at large values of $r,$ we have that when $\Lambda$ is small enough, the surface gravity and the area radius function of the cosmological horizon are given approximately by  
\begin{equation}
\kappa_\Lambda \sim \sqrt{\Lambda/3},\;\; r_{\Lambda}=f_{\Lambda}(\epsilon_i) \sim \sqrt{ 3/\Lambda}, \;\;  for \;\; \Lambda \approx 0^+.
\end{equation}
In fact, at leading order, a sufficiently regular function $U$ in the a neighborhood  of  $\Lambda=0$ satisfying the specifications admits the approximation 
\begin{equation}
U=[F|_{(\Lambda=0)}-\frac{1}{3}\Lambda r^2+ \mathcal{O}(\Lambda^2)]\exp[\psi|_{(\Lambda=0)} + \Lambda h(r) +\mathcal{O}(\Lambda^2)],
\end{equation}
where $h(r)$ is a real function which goes to zero sufficiently rapidly at infinity. Since necessarily $\rho=(2 G \mathcal{A}_{\mathrm{bh}})^{-1} c^{4}$ on $\Upsilon,$ we find accordingly 
\begin{equation}
e^{ (\kappa_- - \kappa_{\Lambda})\upsilon} \approx \exp\{ \sqrt{ \Lambda/3}  [ \; e^{\psi}(p_\theta/\rho)^{1/2} -1  ] \upsilon \} \label{lestimates}
\end{equation}
for the nearly extreme asymptotically de Sitter black holes in the neighborhood of 
$\Upsilon \cap\{ \Lambda=0; p_{\theta} > 0 \}$,  where
\begin{equation}
e^{\psi}=\exp[-(G/c^4)\int^{\infty}_{\mathcal{A}_{\mathrm{bh}}}(2F)^{-1}(\rho+p_r)d\mathcal{A}]. 
\end{equation}
We have restored the appearance of the gravitational constant in the last formulae. And there it is, exactly how the sign and value of the geometrical quantities $\kappa_--\kappa_+$ and $\kappa_--\kappa_\Lambda$; and thus the smallest rate of divergence of the flux of radiation on the inner Cauchy horizon, is determined purely from the knowledge of the stress-energy content of the spacetime: via equations (\ref{rate}), (\ref{lcriterion}) and (\ref{lestimates}) when $\Lambda$ is small enough. 

\textit{It is concluded that if $U\in C^3,$ $U \sim 1-\frac{2GM}{c^2r}-\frac{\Lambda r^2}{3}$ at infinite, $U_{,\epsilon_i}\neq 0$ and $8\pi G p_{\theta}\neq\Lambda c^4$ at a point $x_o=(r_o,{\epsilon_o}_i)\in \Upsilon;$ then, for nearly extreme static nonrotating black holes in the vicinity of ${\epsilon_o}_i$:} 
 
\indent  (i) \textit{The blueshift instability criteria $\kappa_--\kappa_+>0$ implies $\Gamma_{\|}>-[1-(8\pi G p_{\theta})^{-1}\Lambda c^4]$ at $x_o\in \Upsilon,$ and vice versa.}

\indent (ii) \textit{If in addition, $\Lambda$ is a sufficiently small positive constant; then, the necessary condition for inner Cauchy horizon (blueshift) stability $\kappa_--\kappa_\Lambda \leqslant 0$ implies  $e^{2\psi} p_{\theta}  \leqslant \rho$ on $x_o \in \Upsilon,$ and vice versa.}
 
 For nearly extreme nonrotating black holes, the surface gravity estimates depend very heavily on the value and rate of change of the tangential pressure.
 
\section{\label{counterexample}From blueshift to redshift}
 What kind of matter breaks strong cosmic censorship? How do the terms `physically reasonable matter' and `appropriate energy conditions' balance each other? 
 
 In this section we shall show that even invoking the dominant energy condition (that to any observer the local energy density appears nonegative and the local energy flow vector is nonspacelike \cite{Ellis}), it is still possible to curve  the spacetime in such a way as to produce a non-rotating spherically symmetric black hole with an inner Cauchy horizon whose surface gravity is strictly less than the surface gravities of the event and cosmological horizons.   
 
  Let us set the radial pressure $p_r$ equal to the negative of the energy density $\rho,$ so that, according to (\ref{reg}), $\psi$ vanishes, as occurs  with the case of electrodynamics. Hence, $\Gamma_{\bot}$ is zero and the entire spacetime geometry depends solely on the magnitude of $m c^2,$ where the mass function $m$ of equation (\ref{mass}) already takes into consideration the global contribution from the negative gravitational potential energy.  
  
   For  a given constant $b$ consider the equation of state
 \begin{equation}
p_\theta=b^2\rho^2/(e^{b^2\rho}-1) \label{Planck}.
\end{equation}
  Then, the dominant energy condition (which is equivalent to the statement that in any orthonormal basis the energy is nonegative and dominates the other components of $T_{\mu \nu},$ in our case  $|p_j|\leq\rho,$  $j=1,2,3;$ see \cite{Ellis, WaldGR}) is readily satisfied. Such an equation of state may be found in Maxwell and  Born-Infeld  electrodynamics; i.e., $p_{\theta}=\rho$ and $p_{\theta}=\rho (1+ b^2 \rho)^{-1}$ respectively \cite{Born-Infeld}, not only when $b$ goes to zero but also at `low' densities. Only at `ultrahigh' densities, where the properties of matter have not been tested with  great  accuracy, will it exhibit unfamiliar behavior: 
\begin{equation} 
\Gamma_{\|}|_{\rho \rightarrow +\infty} = -\infty;
\end{equation}
compare with $\Gamma_{\|}|_{\rho \rightarrow 0^+} = 2$ in the low energy density regime. Thus, according to (\ref{rate}) and (\ref{lcriterion}),  it might lead to a situation where $\kappa_{-}-\kappa_{+}$ is negative. Moreover,  the tangential pressure $p_{\theta}$ is positive and bounded, strictly less than $\rho$ at ultrahigh densities; therefore, (\ref{Planck}) might  give as well a negative value for $\kappa_{-}-\kappa_{\Lambda},$ see  equation (\ref{lestimates}). Such properties suggest that the matter field here considered might form Cauchy horizons that fail to be surfaces of infinite gravitational blueshift when $\Lambda$ is small enough. To add support to this view  we shall show (Sec.\ref{subsec:find} and Sec.\ref{subsec:curv}) that  
 \begin{equation}
\mathcal{C} \equiv \Upsilon \cap \left \{ (r, \overrightarrow{\epsilon})   | \Gamma_{\|}+1<0 \right \} \label{Cset}
 \end{equation}
is not the empty set and that a curvature singularity exist at $r=0$ inside the black hole. It is worth emphasized that theories of nonlinear electrodynamics obey a Birkhoff's theorem, as well as, the zero and first law of black-hole mechanics \cite{Rashdeed}.     
\subsection{\label{subsec:find}Finding the event and anti-event horizons}
 Since $\rho+p_r$ vanishes identically, the system of equations (\ref{Erho})-(\ref{Gpressure}) reduces to  
\begin{equation}
\rho+\Lambda c^4=\frac{dmc^2}{d\mathcal{V}}, \;\;\  p_\theta-\Lambda c^4=-\frac{d(\mathcal{A}\frac{dmc^2}{d\mathcal{V}})}{d\mathcal{A}}. \label{Gsimple}
\end{equation}
Therefore, 
\begin{equation}
p_\theta d\mathcal{A}+ d(\mathcal{A}\rho)=0 \label{bian}
\end{equation}
not only on $\Upsilon$ but also at any non extreme point of the physical parameter space of the corresponding static hole. 
The forgoing formula (\ref{bian}) which is independent of $\Lambda$, using  (\ref{Planck}), can be regarded as an ordinary differential equation for $\rho.$ By inspection, if $p_\theta$ is negligible, $\rho$ is inversely proportional to $\mathcal{A};$ whereas if $p_\theta \approx\rho,$ the energy density becomes inversely proportional to $\mathcal{A}^2.$  

In order to express the solution in analytic form let us first introduce the function:
\begin{equation}
 \mathcal{S}(u)\equiv u  \exp[\int^{\infty}_{u}dz/(e^z+z-1)].
\end{equation}
\begin{figure}[h]
\includegraphics[width=3.3in]{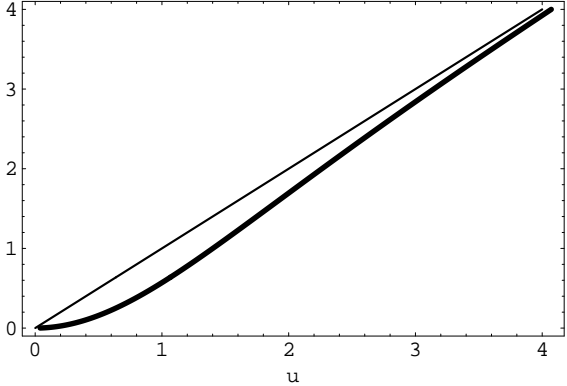} 
\caption{\label{graph} graph of $\mathcal{S}^{(-1)}: \Re^+ \rightarrow \Re^+.$ The thick s-shaped curve refers to the graph of $\mathcal{S}^{(-1)}(u),$ it tends to the identity function at positive infinite values of $u$.}
\end{figure}

which not only is well behaved in the interval $(0,+\infty)\ni u$ but also has an inverse, hereby  denote $\mathcal{S}^{(-1)},$ with the following asymptotic behavior: $\mathcal{S}^{(-1)}(u) \approx u^2$  for $u  \approx +0$, $\mathcal{S}^{(-1)}(u) \approx u$  for $u \approx +\infty.$ The graph of $\mathcal{S}^{(-1)}(u)$ has been depicted in Fig.\ref{graph}.
 
The solution of the  differential equation is then  
\begin{equation}
\rho=  b^{-2}\mathcal{S}^{(-1)}(q r^{-2}), 
\end{equation}  
where $q$ is an integration constant, which we set equal to 
\begin{equation}
q=\sqrt{\varepsilon_o/2} (b|Q|/4\pi \varepsilon_o)
\end{equation}
so that in the limit when $b$ goes to zero the solution tends to the one given by Maxwell's theory; $\varepsilon_o$ denotes the permittivity of free space and $Q$ can be interpreted as the total charge of the hole. 
It is also manifest that $\rho \sim const.  \; r^{-4}$ for $r \approx +\infty$ and $\rho \sim const. \; r^{-2}$ for $r \approx 0^+.$

Directly integrating (\ref{Gsimple}), we obtain the mass function  
\begin{eqnarray}
\frac{2Gm(r)}{c^2}&=& \frac{2GM}{c^2}+\frac{\Lambda r^3}{3}  \label{sol}  \\
&&-\frac{8 \pi G}{b^2 c^4}\int^{\infty}_{r}\mathcal{S}^{(-1)}(\sqrt{\frac{\varepsilon_o}{2}}\frac{b|Q|}{4\pi \varepsilon_o x^2})x^2 dx, \nonumber
\end{eqnarray}
where $M$ is  an integration constant equal to the total mass of the hole (compare this result with the mass for a stationary spherically symmetric Born-Infeld black-hole, see \cite{Gibbons}). The metric is given by
\begin{eqnarray}
ds^2&=&-\left(1-\frac{2Gm(r)}{c^2r}\right)dv^2 +2dvdr+r^2d\Omega^2. \label{line}
\end{eqnarray}
\subsubsection{\label{subsec:neig}The neighborhood of $\Lambda=0$}
 The discovery of the accelerated expansion of the universe \cite{Riess} has given a remarkable observational evidence  in favor of the existence of a small positive cosmological constant pervading the cosmos. The work done in Sec.\ref{subsec:est} allow us to infer some conclusions about near extreme black holes states having a sufficiently  small positive cosmological constant, by restricting our attention to asymptotically flat scenarios. 
 
 In view of this, let us set $\Lambda_o=0$ and see what can be deduced in the  case of the nonlinear electromagnetic theory of Eq.(\ref{Planck}). Since $\rho$ and $p_{\theta}$ are nonegative, according to  (\ref{Gsimple}), the mass function is  a concave, nondecreasing function of the $r$-coordinate. Therefore, no more than two horizons can exist for a vanishing value of $\Lambda.$ 
 
  To find $\Upsilon$ with $\Lambda_o=0$  (points of the present theory representing stationary, spherically symmetric, asymptotically flat, extreme black holes), we need to find simultaneous solutions for the two conditions given by (\ref{extremeSET}). As we have stated previously, this implies that the ${\epsilon_o}_i$'s are not independent. After a little algebra, it can be seen that the corresponding  functional relationship between $Q$ and $M$ can be obtained by solving  
\begin{equation}
\mathcal{S}^{(-1)}(u)=  \sqrt{\varepsilon_o/2} (bc^4/G|Q|) u; \;\;\; u= qr^{-2} \label{xxx}
\end{equation}
This implies finding the intersection of the curve $\mathcal{S}^{(-1)}(u)$ with a straight line of slope $\sqrt{\varepsilon_o/2} (bc^4/G|Q|)$ passing through the origin. 
 
 Let us call $u_c(Q)$ the solutions of this equation. Then according to Fig.\ref{graph}, a solution to (\ref{xxx}) exists if and only if the slope of this straight line lies between one and zero, or equivalently, if and only if 
\begin{equation} 
Q_{\min}^2\equiv 2^{-1}\varepsilon_oc^2  (bc^3/G)^2 < Q^2< +\infty.
\end{equation} 
Moreover,   
\begin{equation}
\lim_{Q^2\rightarrow Q_{\min}^2}u_c(Q)=+\infty  \;\;\; and \;\;  \lim_{Q^2\rightarrow +\infty}u_c(Q)=0^{+}.
\end{equation}
Since $u_c(Q)$ is a continuous function of $Q,$ it follows that $r_o^2=q_ou_o^{-1}=\sqrt{\varepsilon_o/2}(b|Q|/4\pi \varepsilon_o u_c);$ i.e the square of the area radius function of the event horizon of an extreme black hole spans the whole range of positive real numbers $\Re^+$\textemdash i.e. $r_o\in(0,+\infty).$ In particular, it includes the region at ultrahigh densities and small values of  $r_o,$  where $\Gamma_{\|}|_{\rho \rightarrow +\infty} = -\infty.$ There is no obstacle to find $M_c(Q)$ since it appears linearly in (\ref{sol}). The positivity of $M_c(Q)$ is guarantee by the positive mass theorem of general relativity \cite{Yau, Witten1981b, SorkinPE}. Thus $\mathcal{C},$ given by (\ref{Cset}), is not empty.
  
\subsection{\label{subsec:curv}Curvature invariants}
Let us now estimate the curvature invariants: the Ricci scalar, the square of the Riemman tensor, and the Kretschmann invariant in the ultra-high density regime. 

For the present case, we have:
\begin{equation}
R=-c^{-4}(T-4\Lambda c^4)\approx 2c^{-4} \rho,   
\end{equation}
\begin{equation}
R^{\alpha \beta}R_{\alpha \beta}=c^{-8}[T^{\alpha \beta}T_{\alpha \beta}-2\Lambda c^4 (T-2\Lambda c^4)]\approx 2 c^{-8} \rho^2,
\end{equation}
and
\begin{eqnarray}
R^{\alpha \beta \gamma \delta}R_{\alpha \beta \gamma \delta}&=&{\;^{(2)}R}^2+8r^{-2}r^{|a b}r_{|a b}+4r^{-4}(1-r^{,a}r_{,b})^2 \nonumber \\ &\approx& 4[(\mathcal{E}/4\pi r^3)^2+ 2(\mathcal{E}/4\pi r^3 + \rho)^2],  
\end{eqnarray}
since the energy density dominates over the tangential pressure. We have used $T$  for the trace $g^{\alpha \beta}T_{\alpha \beta},$ $\;^{(2)}R$ for the Ricci scalar of the two-dimensional metric $g_{ab}$ of (\ref{smetric}), the symbol $(|)$ for its associated covariant derivative, and $\mathcal{E}$ for
\begin{equation}
\mathcal{E}=(4 \pi/b^2)\int^{\infty}_{0}\mathcal{S}^{(-1)}(\sqrt{\frac{\varepsilon_o}{2}}\frac{b|Q|}{4\pi \varepsilon_o x^2})x^2 dx -Mc^2; 
\end{equation}
 the last quantity having the interpretation of the binding energy. But according to equation (\ref{Bianchi}), $\rho \sim const. \; \mathcal{A}^{-1}$.  Therefore, at $r=0$ there is a scalar polynomial curvature singularity which incidentally is free of pressure, compare with Fig.\ref{wormhole}(a).

For vanishing $\Lambda,$ the existence of two horizons can be guarantee in the interval
\begin{equation} 
M_c(Q)<M<\frac{4 \pi }{b^2c^2}\int^{\infty}_{0}\mathcal{S}^{(-1)}(\sqrt{\frac{\varepsilon_o}{2}}\frac{b|Q|}{4\pi \varepsilon_o x^2})x^2 dx<\infty
\end{equation} 
for $Q>Q_{\min},$ i.e., when $\mathcal{E}$ is positive. By continuity, the addition of a sufficient small, nonegative cosmological constant $\delta \epsilon_{\Lambda}=\Lambda$ can only produce a tiny shift from the original values of the horizon's positions; c.f. Eq.(\ref{area}). 
 
The near extreme black holes derived from (\ref{sol}) have a causal structure similar to the one obtained for a Reissner-Nordstr\"{o}m--de Sitter black hole, see Fig \ref{wormhole}(a); however, the blueshift stability issue is another matter: In contrast to the Reissner-Nordstr\"{o}m--de Sitter case, the stability criteria implied by (i) and (ii) of Sec.\ref{sec:reason} are satisfied by `small' black holes of the nonlinear electromagnetic theory given by (\ref{Planck}).  
\section{Concluding remarks} 

  The strong cosmic censorship hypothesis asserts roughly that: ``Every generic inextendible spacetime $\mathcal{M}$  which evolves, according to classical general relativity, with matter that satisfies appropriated energy conditions and physically reasonable equations of state, from generic non-singular initial data on a complete spacelike hypersurface $\Sigma,$ is globally hyperbolic."  This implies that  $\mathcal{M}$ has no Cauchy horizon, and in fact, by a theorem of Geroch \cite{Geroch}, that $\mathcal{M}$ is homeomorphic to $\Re \times \Sigma.$ The strong cosmic censorship question then becomes a stability problem of Cauchy horizons.  
  
 If one believes in the veracity of the conjecture, it should be possible to find a set of generic conditions giving a precise meaning to the terms ``appropriated energy conditions" and ``physically reasonable equations of state" that imply the instability of Cauchy horizons; in particular, the ones hidden inside black holes solutions.  
   
With all reserve, the analysis of Sec.\ref{sec:reason} invite us to suggest, tentatively, the following limited version of the strong cosmic censorship  hypothesis for black hole interiors:

\textit{$\Gamma-$version of the SCC conjecture: 
Every generic four-dimensional spacetime $\mathcal{M}$ containing a black hole with regular event horizon and satisfying the following four conditions, has a maximal future development which is locally inextendible in a $C^3$ manner inside the hole:} \\
\indent (a) \textit{$\mathcal{M}$ arises from the evolution of nonsingular initial data given in a complete spacelike hypersurface $\Sigma$,} \\
\indent (b) \textit{the initial data evolves in a $C^3$ manner according to classical general relativity,} \\
\indent (c) \textit{at late times the corresponding hole asymptotically approaches to a `nearly' extreme spherically symmetric static configuration,} \\
\indent (d) \textit{at the event horizon of static spherically symmetric black holes with zero surface gravity, the matter sources are such that to satisfy the following inequality $\Gamma_{\|} > -[1-( 8\pi G p_{\theta})^{-1}\Lambda c^4],$ where $\Gamma_{\|}$ is the adiabatic index given by  $\Gamma_{\|}\equiv [1+ (\rho/p_\theta)](\partial p_\theta/ \partial\rho)_S$.}

By future development we mean the collection of all the points $x \in \mathcal{M}$ with the property that each past directed causal curve $\gamma$ without past end point containing $x$ intersects $\Sigma$ at one and only one point. The inclusion of the $C^3$ requirement on the metric is related with the fact that the arguments presented here involve the use of the Bianchi identity and the local conservation of material energy over a small region of the spacetime. Perhaps this condition could be refined somewhat. However,  Dafermos \cite{Dafermos} has shown that under the restriction of spherical symmetric the Einstein-Maxwell-scalar system has solutions where  a continuos extension of the metric beyond the maximal future development is possible but where a $C^1$ extension fails.   
  
  According to linear perturbation theory, the matter condition  $\Gamma_{\|} > -[1-( 8\pi G p_{\theta})^{-1}\Lambda c^4]$ at  the event horizon of a static spherically symmetric black hole $X,$ with zero surface gravity,  signifies that an observer who attempts to travel into new worlds through the interior of black-hole configurations sufficiently close to X  will detect a shell of infinitely intense radiation at the instant of his crossing the Cauchy horizon (if any). This type of (blueshift) instability, where the radiation  gets compress to an infinite energy density, suggest that a curvature singularity may develop, sealing the region where predictability is lost, see Fig.\ref{wormhole}; however, higher-order nonlinear terms must be included to know  how  the internal structure of the hole is really affected. Thus, the conjectural nature of the above (presumably sufficient) conditions for the validity of the strong cosmic censorship principle inside black holes.  

In this regard, a very reasonable picture called the `mass inflation scenario' \cite{PoissonIsrael} has emerged  from the study of some physically relevant systems, leaving a globally hyperbolic spacetime as the end result of gravitational collapse. Mass inflation is a nonlinear phenomenon  where the radiation escaping from the surface of the collapsing object, say a star, couples with any incident flow of the incoming radiation, leading to an unbounded increase in the effective gravitational mass parameter of the hole. The space-time curvature blows up at or before one reaches the Cauchy horizon. This raises the question of whether there is  a clear mathematical connection between the blueshift instability and the occurrence of mass inflation, a nonlinear problem lying beyond the scope of the present paper. 
 
   Numerical and analytical studies of gravitational collapse for the Maxwell-Einstein system \cite{Dafermos} as well as for self-gravitating charged scalar fields \cite{Hod} indicate the formation of a null, gravitationally weak, mass-inflation singularity along the Cauchy horizon, which is a precursor of a strong, spacelike singularity along the geometric place $r=0$ \cite{Frolov}. If the hole is supplied with a spin, the curvature scalar $R^{\alpha \beta}R_{\alpha \beta}$ oscillates infinitely many times, but not in a chaotic way, while diverging in the proximity of the null, gravitationally weak singularity at the Cauchy horizon \cite{Ori}. It may be that in physically realistic situations, this is the general picture, but we do not yet know. 

  We have replaced the common phrase ``the matter sources are such that to satisfy the $X$ energy condition and physically reasonable equations of state, say Y" with the condition $\Gamma_{\|} > -[1-( 8\pi G p_{\theta})^{-1}\Lambda c^4]$ at the black hole event horizon, where the expression in square brackets is nonegative. In doing so we have placed the theory into a form which resembles those statements from the stability theory of relativistic stars, where instability criteria can be given in terms of inequalities containing the adiabatic index \cite{Lindblom, Shapiro, Stellar, Sri, MTW}.   
 
 We see, for instance, that the above condition is satisfied if  $X$ is replaced by the weak energy condition and $Y$ is chosen to be $(\partial p_\theta / \partial\rho)_S>0,$ with a  positive tangential pressure $p_\theta>(\Lambda c^4/8\pi G)$ near the black-hole horizon. The weak energy condition is the assumption that the local energy density as measured by any observer is nonnegative.
 
 Without modification, the present analysis does not allow us to infer a proposal about the very special case when  $p_\theta=(\Lambda c^4/8\pi G)$  at the horizon of an static extreme hole, since this would imply that $U''$ also vanishes in such a place; hence, the locus of the zeroes of the function $U,$ i.e., the positions of the event and anti-event horizons, can not be deduced from (\ref{area}). Nevertheless, a generalisation of the main procedure seems feasible (under the assumption of local existence of only two solutions, $r_-=f_-(\epsilon_i)$ and $r_+=f_+(\epsilon_i),$ in the neighborhood of an extreme state) taking  higher order terms in the expansion (\ref{expansion}) unless $p_{\theta}-(\Lambda c^4/8\pi G)$ vanishes identically. To obtain a reasonable statement that goes beyond near extreme states requires the introduction of global methods.  
   
  The forgoing (blueshift) instability condition relies very heavily on the value and gradient of the tangential pressure $p_\theta$. Thus, the question arises of how  this relation gets modified when the unperturbed metric corresponds to a rotating hole. The formal extension of this analysis to the rotating case is in progress. In this connection, it is worth noting that regularity   conditions (and thus absence of curvature singularities) imposed on any horizon highly constrain the near horizon spacetime geometry, and thus the form of the stress energy tensor. In the special case of spherical symmetry, it can be seen from (\ref{reg}) that $\rho + p_r$ vanishes at the horizon whenever this is regular. Similar relations hold in more complex circumstances \cite{dirty}.   
 
 In Sec.\ref{counterexample}, we provide evidence suggesting that a tiny cosmological constant and the positivity of tangential pressures can lead, in some special situations, to the blueshift stability of Cauchy horizons hidden inside black holes, even when the dominant energy condition is satisfied. However, this is at the cost of admitting  a domain where velocity of sound $(\partial p_{\theta}/ \partial \rho)_S^{1/2}$ is an imaginary number, or more generally, when without the help of gravity Le Chatelier's principle is violated. This principle is usually stated as follows,  
         
 \textit{Le Chatelier's principle \cite{Chatelier}: Every change of one of the factors of a physical equilibrium occasions a rearrangement of the system in such a direction that the factor in question experiences a change in a sense opposite to the original change."}
   
 The particular example of Sec.\ref{counterexample} is a modification of Maxwell electrodynamics and the violation of Le Chatelier's  principle occurs only in the realm of ultrahigh densities due to an exponential fall off of the tangential pressure as a function of the energy density; thus, implying the existence of an open set of states in \textit{unstable} equilibrium that can be connected, by a continuos sequence of equilibrium configurations, to an open set of states in \textit{stable} equilibrium. At low energy densities, the equation of state of the theory, see Eq.(\ref{Planck}), reduces to the one derivable from the Born-Infeld Lagrangian \cite{Born-Infeld}: 
 $$\mathcal{L}=b^{-2}\{\sqrt{-g}-\sqrt{-\det(g_{\mu \nu} + b F_{\mu \nu})}\}$$ 
which arises in open superstring theory, where $F_{\mu \nu}$ is the Maxwell strength tensor. Note that by  combining  $\rho=-\mathcal{L}+E\cdot \partial \mathcal{L}/\partial{E}$ with $p=\mathcal{L}-B\cdot \partial{\mathcal{L}}/\partial B,$ where $E$ and $B$ are the electric and magnetic fields, we obtain $p=\rho(1+b^2 \rho)^{-1}.$  In 1934, Born and Infeld suggested to put forward the classical foundations of electrodynamics on a similar basis than the ones of  relativistic mechanics: To place an upper limit to the magnitude of purely electric  fields, just like the velocity of light  is an upper limit to the velocity of massive particles; to use as a Lagrangian of the theory the square root of a determinant (area), just like the Lagrangian of massive particles is the square root of a determinant (proper length); Maxwell's theory is recovered in the limit where $b^{-1}$ goes to infinity, just like Newtonian mechanics is recovered from special relativity in the limit where the velocity of light goes to infinity. Instead, (\ref{Planck}) is inspired on Planck's radiation formula which led  to the discovery of quantum mechanics in 1900: $$\varrho(\nu,T)=(8\pi h \nu^3/c^3)[\exp(h\nu/kT)-1]^{-1},$$ where $\varrho$ is the spectral density of blackbody radiation in thermal equilibrium as a   function of the frequency  $\nu$ and the temperature $T,$ $k$ is Boltzmann's constant, and $h$ is Planck's constant.  
  
  The requirement $(\partial p/ \partial \rho)_S \geq 0$  has been used as an assumption in calculations of the maximum mass of the equilibrium configuration of neutron stars when the equation of state of matter, modelled as a perfect fluid, is unknown \cite{Ruffini}.   It provides, in this case, a microscopic stability condition avoiding the local spontaneous collapse of matter. 
  
 The example given in section \ref{counterexample} illustrates how collapsing matter, that is capable of producing singular solutions of the matter field equations in a fixed flat spacetime could lead to a violation of strong cosmic censorship, see also \cite{Penrose1979, Wald97}. Perhaps there is a profound connection between Le Chatelier's principle, in its various forms, and the fundamental question of cosmic censorship.

  Another line of investigation with implications for the present formulation  of the cosmic censorship principle is the study of the existence/non-existence of solutions to the equations of geodesic motion describing outgoing nonspacelike curves terminating at a \textit{singular} point in the past \cite{Joshi}. 
The main observation, as far as the appearance of globally naked singularities is concerned, is that under the restriction of spherical symmetry the apparent horizon is a sub-solution of the radial null geodesic equation provided the weak energy condition and the inequality $p_r\geq \max \{-\rho, -(8 \pi G)^{-1}r^{-2}c^4 \}$ on the radial pressure are fulfilled \cite{Giambo}. This allows the application of comparison techniques of ordinary differential equations leading to the determination of the spectrum of end states, black holes or globally naked singularities, of spherical gravitational collapse. 

Attention is reserved mostly to shell focusing singularities, since the other type of singularities, shell crossing singularities, are `thought' to be removable by a possible extension of the spacetime \cite{Joshi}. 

By invoking regularity at the center prior the  singularity formation and using a system of coordinates due to Ori \cite{Oricoo}, the picture obtained is one where the end result of collapse is selected according to the value of a positive integer $n$: the equation of state, the value of the cosmological constant, the initial profiles for density, pressure, velocity, and so on, play the role of mere inputs for the so called ``n-machine"  unlesss $n$ becomes three, in which case  they combine to  form  a critical non-dimensional parameter $\xi$ ruling the transition between a black hole or a globally naked central singularity \cite{Giambo2}. 
 
 Now, one might have locally or globally naked singularities arising from regular initial data and satifying  the usual energy conditions \cite{Joshi, Frolov}. But the existence of a naked singularity is not enough to provide a counterexample for cosmic censorship. A stability test is also required under physically reasonable perturbations, say against the addition of different classes of matter fields or gravitational waves in every possible state of excitation, and this is when the technical problems commence.  The problem of defining a global nonlinear stability theory for Einstein's field equations remains a challenged for the future, and so the stability analysis of the above spectrum of states.

To conclude, the relevance of cosmic censorship to the physics of black holes  is perfectly illustrated by Chandrasekhar's remarks  at the opening of his treatise \textit{The Mathematical Theory of Black Holes} \cite{Chandra}: 
  
  ``The black holes of nature are the most perfect macroscopic objects there are in the universe; the only elements in  their construction are our concepts of space and time. And since general theory of relativity provides only a single unique family of solutions for their descriptions, they are the simplest objects as well."
 
  Indeed, secretly hidden behind the wisdom and beauty of Chandrasekhar's words is the hypothesis of cosmic censorship: in its weak version, it deals with the  generic macrostability of black holes necessary for their own physical existence; in the strong variant it protect us, classically, from a spacetime geometry that is not unique. Cosmic censorship violation would certainly spoil black hole's simplicity, see Fig.\ref{wormhole}(a). 
  
  The equations shown in this letter  give a first glimpse of the unlikelihood (or likelihood) of the rupture of cosmic censorship as suggested by a special type of stability analysis. The answer to Penrose's riddle, however, is still waiting for discovery, perhaps in an strange but subtle form of mathematical beauty and physical intuition.

\emph{Acknowledgments.} I would like to thank Steven Carlip for useful discussions relative to this problem and for reading a previous version of the manuscript. This research has been sponsored by the \textsc{UC MEXUS-CONACyT} fellowship program under the grant\textendash $040013,$ partial funding is also provided by the US Department of Energy, grant $DE$\textendash $FG02$\textendash $91ER40674$.

\end{document}